\documentclass[seceq,supplement]{ptptex}

\usepackage{graphicx}
\usepackage{wrapft}

\title{%        %You can use \\ for explicit line-break
Exact many-electron ground states \\on diamond and triangle Hubbard
chains }

\author{%       %Use \scshape  for the family name
Zsolt \textsc{Gul\'acsi}$^{1,2}$, Arno \textsc{Kampf}$^{1}$, and
Dieter \textsc{Vollhardt}$^{1}$ }

\inst{%     %Affiliation, neglected when [addenda] or [errata]
$^1$Theoretical Physics III, Center for Electronic Correlations and Magnetism,
Institute of Physics, University of Augsburg, D-86135 Augsburg, Germany \\
$^2$Department of Theoretical Physics, Faculty of Natural Sciences,
University of Debrecen, H-4010 Debrecen, Hungary
}

\abst{We construct exact ground states of interacting electrons on
triangle and diamond Hubbard chains. The construction requires (i) a
rewriting of the Hamiltonian into positive semidefinite form, (ii)
the construction of a many-electron ground state of this
Hamiltonian, and (iii) the proof of the uniqueness of the ground
state. This approach works in any dimension, requires no
integrability of the model, and only demands sufficiently many
microscopic parameters in the Hamiltonian which have to fulfill
certain relations. The scheme is first employed to construct exact
ground state for the diamond Hubbard chain in a magnetic field.
These ground states are found to exhibit a wide range of properties
such as flat-band ferromagnetism and correlation induced metallic,
half-metallic or insulating behavior, which can be tuned by changing
the magnetic flux, local potentials, or electron density. Detailed
proofs of the uniqueness of the ground states are presented. By the
same technique exact ground states are constructed for triangle
Hubbard chains and a one-dimensional periodic Anderson model with
nearest-neighbor hybridization. They permit direct comparison with
results obtained by variational techniques for $f$-electron
ferromagnetism due to a flat band in CeRh$_3$B$_2$. }
\begin{document}

\maketitle

\section{Introduction}

Electronic correlations in solids can result in a high sensitivity
of these systems to small changes in external parameters such as
temperature, pressure, magnetic field, or doping
\cite{Review1,Review2,Review3}. This may then lead to strongly
non-linear responses such as large resistivity and volume changes, a
strong thermoelectric response, colossal magnetoresistance, and
high-$T_c$ superconductivity. The complexity displayed by
electronically correlated materials is not only the topic of
fascinating fundamental research, but is also significant for
technological applications, e.g., in sensors, switches, and in
spintronics.

Even highly simplified models of interacting electrons
are known to display a dazzling diversity of phases,
e.g., metallic, insulating, and many types of long-range ordered
states \cite{Montorsi,Kivelson2007}. Theoretical investigations of
correlation effects almost always make some kind of approximation
which cannot, in general, explain the full complexity of a model in
detail. In this situation exact solutions of correlation models are
particularly important and useful.

In this paper we discuss a general strategy for the construction of
exact many-electron ground states of lattice models such as the
Hubbard model. The approach is based on three distinct steps: (i)
the rewriting of the Hamiltonian in positive semidefinite form,
(ii) the construction of exact ground states for this Hamiltonian,
and (iii) the proof of the uniqueness of the ground states
constructed thereby.

This strategy was recently employed to construct a class of exact
ground states of a three-dimensional periodic Anderson model (PAM),
including the conventional PAM on regular Bravais lattices at and
above 3/4 filling \cite{re12,re13}, and to the diamond Hubbard chain
in a magnetic field \cite{re15}. In both cases a wide range of
features such as metallic, non-Fermi liquid, and insulating phases
in the case of the PAM \cite{re12,re13}, and flat-band
ferromagnetism, correlation-induced metallic, half-metallic, or
insulating behavior in the case of the diamond Hubbard chain
\cite{re15} was found. These exact results prove that the phase
diagram of even seemingly simple quantum mechanical models of
many-particle systems can indeed be surprisingly complex.

The construction of exact ground states by casting the Hamiltonian
into positive semidefinite form was originally applied by Brand and
Giesekus to the Hubbard model and the PAM on special perovskite-type
decorated lattices in dimensions $d\geq2$ at $U=\infty$ \cite{re1}.
The procedure was generalized by Strack  \cite{re3} to the PAM and
the extended Emery model on regular lattices in low dimensions
\cite{re2}, and to the extended Hubbard model. The method was
subsequently used to derive rigorous criteria for the stability of
saturated ferromagnetism in general Hubbard-type models \cite{re4}.
The possibility of superconductivity in the extended Hubbard model
and its strong coupling limit was also investigated
\cite{sup1,sup2,sup4}. Furthermore, next-nearest neighbor terms in
the Hamiltonian were shown to cause charge density waves, degenerate
non-magnetic localized phases, and phase separation \cite{re7}.
Introducing operators to rewrite the Hamiltonian in positive
semidefinite form which are defined on the entire unit cell of a
$d$-dimensional lattice (e.g., on plaquettes in $d=2$ or cubes in
$d=3$) rather than only on bonds,  Orlik and Gul\'acsi \cite{re5}
investigated the PAM on regular two- and three-dimensional lattices
at $U=\infty$, finding insulating non-magnetic phases. This method
was used by Sarasua and Continentino \cite{re10} to analyze the
one-dimensional PAM in more detail. Gurin and Gul\'acsi \cite{re8}
and Gul\'acsi \cite{re9} demonstrated that it is also possible to
investigate the PAM in $d=2,3$ at $U < \infty$ . The exact ground
states constructed by Gul\'acsi and Vollhardt for the PAM at $3/4$
\cite{re12} and $1/4$ \cite{re13} gave the first mathematical proof
for the presence of ferromagnetism in this model in dimension $d=3$.
These authors also detected a strong anomaly in the compressibility
of the system at the boundary of the parameter region where the
non-magnetic localized phase is stable --- a feature which is
typical for $f$ electron systems such as Cerium. In this connection
the uniqueness of the constructed states was proved explicitly for
the first time.

Most recently, the present authors constructed exact ground states
of interacting electrons on the diamond Hubbard chain in a magnetic
field \cite{re15}.  This model was shown to exhibit a wide range of
properties such as flat-band ferromagnetism and correlation induced
metallic, half-metallic or insulating behavior which can be tuned by
changing the magnetic flux, local potentials, or the electron density.

In this paper we present details of the construction of exact
many-electron ground states on the diamond Hubbard chain and of their
physical properties \cite{re15}. In particular, in the Appendices A
and B we provide details of the mathematical proof of the uniqueness
of the ground-state solutions for this model.
We also construct exact ground states on triangle Hubard chains and
show that within a particular subspace of parameters the triangle
Hubbard chain may be mapped to a one-dimensional PAM. The exact
solutions obtained for this model can be applied to investigate the
correlated electron material CeRh$_3$B$_2$, for which Kono and
Kuramoto \cite{Dia10} had constructed an approximate solution.

The paper is structured as follows: In Sec. 2 we discuss the general
strategy for the construction of exact many-electron ground states.
Exact ground states are constructed for the diamond Hubbard chain in
Sec. 3. Details of the proof of the uniqueness of these ground
states are presented in the appendices A and B.  The technique is
employed in Sec. 4 to obtain exact ground states also for the
triangle Hubbard chain, and the results are applied to explain
properties of CeRh$_3$B$_2$.

\section{Construction of exact many-electron ground states: General Strategy}

Our construction of exact many-electron ground states
is based on the following three steps.

\subsection{Step 1: Rewriting the Hamiltonian in positive semidefinite form}

In the first step the many-electron Hamiltonian is cast into
positive semidefinite form. This means that one rewrites the
Hamiltonian in terms of a number of positive semidefinite operators
$\hat P_n$ as
\begin{eqnarray}
\hat H = \hat H_0 + \hat H_U = \sum_{n=1}^L \hat P_n + E_g \equiv
\hat H' + E_g,
\label{eq1}
\end{eqnarray}
where $E_g$ is the ground-state energy. Here (\ref{eq1}) is supposed
to be an operator identity, involving no approximations. The
positive semidefinite nature of $\hat P_n$ is expressed by the
relation $\langle \Psi | \hat P_n | \Psi \rangle \geq 0$, which must
be satisfied for all $|\Psi\rangle$. Due to this inequality, the
minimal eigenvalue of $\hat P_n$ is zero. This property, i.e., the
existence of a well-defined lower bound, namely zero, in the
spectrum of $\hat H'=\hat H - E_g$ in (\ref{eq1}) is the corner
stone of the procedure. Simple examples for a positive semidefinite
operator are $\hat P_n = \hat \Omega^{\dagger} \hat \Omega$, or
$\hat P_n = \hat \Omega \hat \Omega^{\dagger}$, or $\hat P =
\sum_{\bf i} \hat P_{\bf i}$, where the positive semidefinite
operator $\hat P_{\bf i} = \hat n_{{\bf i},\uparrow} \hat n_{{\bf
i},\downarrow} - (\hat n_{{\bf i},\uparrow} + \hat n_{{\bf
i},\downarrow}) +1$ assumes its minimum eigenvalue if at least one
electron is present on the site $\bf i$.

The scalar $E_g$ in (\ref{eq1}), i.e., the ground state energy, is
usually a complicated function of the parameters entering in $\hat
H$. It should be noted  that the decomposition of $\hat H$ in
positive semidefinite form is, in general, not unique and can be
obtained in different ways. Each transformation is valid only in
certain parts ${{\cal{R}}_p}$ of the parameter space.

\subsection{Step 2: Construction of the exact many-electron ground states}

After the transformation of a given Hamiltonian $\hat H$ into a
particular positive semidefinite form valid in a parameter space
region ${{\cal{R}}_p}$,  the exact ground state of $\hat H$ needs to
be constructed. Since $\hat H'=\sum_{n=1}^L \hat P_n$, and the
minimal eigenvalue of each $\hat P_n$ is zero, this requires the
construction of the most general state $|\Psi_g\rangle$ which
satisfies
\begin{eqnarray}
\hat P_n |\Psi_g \rangle = 0,
\label{eq2}
\end{eqnarray}
 for all $n$. Consequently,
\begin{eqnarray}
\hat H |\Psi_g \rangle = E_g |\Psi_g \rangle
\label{eq3}
\end{eqnarray}
holds, where $|\Psi_g \rangle$ is the ground state, with $E_g$ as
the ground state energy in ${{\cal{R}}_p}$. The actual construction
of $|\Psi_g \rangle$ depends on the explicit expression of the
operators $\hat P_n$ in (\ref{eq1}). Therefore each case must be
individually analyzed. Nevertheless the following guideline may be
applied (here we restrict ourselves to fermionic particles):

i) For operators $\hat P_n = \hat \Omega^{\dagger} \hat \Omega$ the
ground state has, in general, the form $|\phi_g\rangle = \hat
O^{\dagger}|0\rangle$, where $|0\rangle$ is the vacuum, and the
anti-commutation relation $\{ \hat \Omega, \hat O^{\dagger} \} =0$
holds. Indeed, in this case one has $\hat P_n |\phi_g\rangle =0$. To
construct the global ground state one then tries to extend this
property to all operators $\hat P_n$.

ii) Likewise, for operators $\hat P_n = \hat \Omega \hat
\Omega^{\dagger}$  the ground state construction makes use of
$(\Omega^{\dagger})^2=0$ and starts from states of the form
$|\psi_g\rangle = \hat \Omega^{\dagger}|0\rangle$, since in this
case $\hat P_n |\psi_g\rangle =0$. To obtain $|\Psi_g \rangle$ from
$|\psi_g\rangle$ the property $\hat P_n |\psi_g\rangle =0$, valid for
a particular $n$, must be extended to all operators $\hat P_n$.

\subsection{Step 3: Proof of the uniqueness of the ground states}

Finally
the constructed ground state has to be shown to be unique, i.e.,
one must prove that there exist no other linearly independent
eigenstates of $\hat H$ in ${{\cal{R}}_p}$ with the same energy
$E_g$. This step employs the property of the \emph{kernel} of $\hat
H'$ which is defined as the Hilbert subspace of states
$|\Phi\rangle$ with $\hat H'|\Phi\rangle =0$, e.g.
\begin{eqnarray}
ker(\hat H'):= \{ \: |\Phi \rangle \: \vert \: \hat H'|\Phi\rangle =
0 \} .
\label{eq4}
\end{eqnarray}
Since $\hat H'=\sum_{n=1}^L \hat P_n$, one has $ker(\hat H') =
\bigcap_{n=1}^L ker(\hat P_n)$. To prove the uniqueness one must
show that $|\Psi_g\rangle$ spans $ker(\hat H')$. For this it is
necessary and sufficient to prove that \\
(i) $|\Psi_g\rangle$ is an element of the intersection of the
kernels of all positive semidefinite operators $\hat P_n$, i.e.,
$|\Psi_g\rangle \in
ker(\hat H')= \bigcap_{n=1}^L ker(\hat P_n)$, and that\\
(ii) all states $|\Psi\rangle \in \bigcap_{n=1}^L ker(\hat P_n)$ can
be written in terms of the constructed ground state
$|\Psi_g\rangle$.

We note that if condition (ii) is not satisfied the constructed
ground state is not unique, implying that other, linearly
independent states with the same lowest energy $E_{g}$ exist. The
uniqueness of the ground states constructed in this paper is proved
in  Appendices A and B.

\subsection{Discussion}

The procedure described above works, in principle, in any dimension
and for any model, independent of the concept of integrability. On
the other hand, a decomposition of $\hat H$ into positive
semidefinite form and a subsequent construction of the ground state
may not always be possible --- especially for Hamiltonians with a
very simple structure. Indeed, the procedure
works the better the more microscopic parameters (e.g., hopping
amplitudes, bands, etc.) enter in $\hat H$. This is due to the fact
that in a high-dimensional parameter space it is easier to find a
subspace ${{\cal{R}}_p}$ where a relation
between microscopic parameters is found, for which the transformation
of the Hamiltonian into positive semidefinite form holds.

\section{Exact many-electron ground states on diamond Hubbard chains}

%%%%%%%%%%%%%%%%%%%%%%%%%%%%%%%%%%%%%%%%%%%%%%%%%%%%%%%%%%%%%%%%%%%%%%%%%
% FIGURE 1.
%%%%%%%%%%%%%%%%%%%%%%%%%%%%%%%%%%%%%%%%%%%%%%%%%%%%%%%%%%%%%%%%%%%%%%%%%
\begin{wrapfigure}{l}{6.6cm}
\centerline{\includegraphics[width=6 cm,height=3 cm]{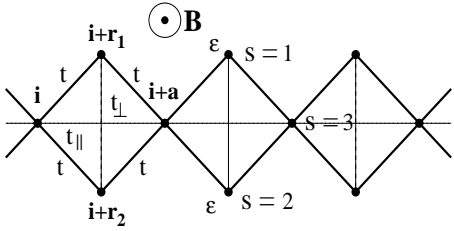}}
\caption{Diamond Hubbard chain; the hopping matrix elements on
different bonds are indicated, as well as the sites on the
sublattices $s=1$ and 2 with potential $\varepsilon$.} \label{fig:1}
\end{wrapfigure}
%%%%%%%%%%%%%%%%%%%%%%%%%%%%%%%%%%%%%%%%%%%%%%%%%%%%%%%%%%%%%%%%%%%%%%%%%

The first exact many-electron ground states for the diamond Hubbard
chain in an external magnetic field have been deduced by Gul\'acsi,
Kampf, and Vollhardt \cite{re15}. Here we provide further details of
the construction process and the physical properties of this system.

The diamond Hubbard chain is shown in Fig. \ref{fig:1}.
This chain  has 3 sites per
unit cell, hence 3 sublattices, denoted by the sublattice index $s=1,2,3$,
and therefore 3 bands. One denotes by $N, N_c$, and $n=N/(3N_c)$ the
number of electrons, cells, and electron density, respectively.
The Hamiltonian of the chain $\hat H=\hat H_0 +\hat H_U$ is given by
\begin{eqnarray}
\hat H_0 &=& \sum_{\sigma} \sum_{{\bf i}=1}^{N_c} \{ [t e^{i \frac{\delta}{2}}
(\hat c^{\dagger}_{{\bf i}+{\bf r}_2,\sigma} \hat c_{{\bf i},\sigma} +
\hat c^{\dagger}_{{\bf i}+{\bf a},\sigma} \hat c_{{\bf i}+{\bf r}_2,\sigma} +
\hat c^{\dagger}_{{\bf i}+{\bf r}_1,\sigma} \hat c_{{\bf i}+{\bf a},\sigma} +
\hat c^{\dagger}_{{\bf i},\sigma} \hat c_{{\bf i}+{\bf r}_1,\sigma})
\nonumber\\
&+&t_{\perp} \hat c^{\dagger}_{{\bf i}+{\bf r}_2,\sigma} \hat c_{{\bf i}+
{\bf r}_1,\sigma}
+ t_{\parallel} \hat c^{\dagger}_{{\bf i}+{\bf a},\sigma}
\hat c_{{\bf i},\sigma} + H. c.]
+ \varepsilon \sum_{s=1,2} \hat n_{{\bf i}+{\bf r}_s,\sigma} \}, \quad \\
\hat H_U &=& U \sum_{{\bf i}=1}^{N_c} \sum_{s=1}^{3} \hat n_{{\bf i}+{\bf r}_s,
\uparrow} \hat n_{{\bf i}+{\bf r}_s,\downarrow}.
\label{dia1}
\end{eqnarray}

%%%%%%%%%%%%%%%%%%%%%%%%%%%%%%%%%%%%%%%%%%%%%%%%%%%%%%%%%%%%%%%%%%%%%%%%%
% FIGURE 2.
%%%%%%%%%%%%%%%%%%%%%%%%%%%%%%%%%%%%%%%%%%%%%%%%%%%%%%%%%%%%%%%%%%%%%%%%%
\begin{wrapfigure}{l}{6.6cm}
\centerline{\includegraphics[width=6 cm,height=3 cm]{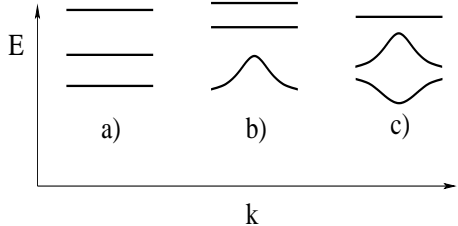}}
\caption{Schematic view of three particular bare band structures of
the diamond Hubbard chain for (a) $t_{\perp}=t_{\parallel}=0$,
$\delta=\pi/2$, (b) $t_{\parallel}
> 0$, ${\varepsilon}=-t_{\perp}+t_{\perp}^{-1}/2$, $\delta=\pi$, and
(c) $t_{\parallel}=0$, $\varepsilon t_{\perp}\cos\delta =t_{\perp}^2
- \cos^2\delta$.} \label{fig:2}
\end{wrapfigure}
%%%%%%%%%%%%%%%%%%%%%%%%%%%%%%%%%%%%%%%%%%%%%%%%%%%%%%%%%%%%%%%%%%%%%%%%%

In the external magnetic field the hopping matrix elements acquire
the Peierls phase factor with $\delta=2\pi\Phi/\Phi_0$ and
$\Phi_0=hc/e$ is the flux quantum. Here we have chosen the vector
potential ${\bf A}\parallel{\bf a}$, and the field dependent hopping
amplitudes follow from $t_{{\bf j},{\bf j}'}( {\bf B})=t_{{\bf
j},{\bf j}'}(0)\exp[(2{\rm i}\pi/\Phi_0)\int_{\bf j}^{{\bf j}'}{\bf
A}\cdot {\rm d} {\bf l}]$. For one flux quantum per unit cell
(triangle) $\delta=\pi$. The Zeeman term is not explicitly included
in $\hat H$, but all deduced ground states have also been analyzed
in the presence of the Zeeman term.

Fourier transforming $\hat H_0$ one finds $\hat H_0 =\sum_{{\bf
k},\sigma} \sum_{s,s'=1}^3M_{s,s'}({\bf k})\hat c^{\dagger}_{s,{\bf
k},\sigma} \hat c_{s',{\bf k},\sigma}$, where the matrix elements
$M_{s,s'}$, which determine the free-electron dispersion, are given
in Ref.\cite{re15}. In the following all energies will be given in
units of $2t$. Examples for bare band structures are schematically
shown in Fig. \ref{fig:2}.

\subsection{Flat-band ferromagnetism}

%%%%%%%%%%%%%%%%%%%%%%%%%%%%%%%%%%%%%%%%%%%%%%%%%%%%%%%%%%%%%%%%%%%%%%%%%
% FIGURE 3.
%%%%%%%%%%%%%%%%%%%%%%%%%%%%%%%%%%%%%%%%%%%%%%%%%%%%%%%%%%%%%%%%%%%%%%%%%
\begin{wrapfigure}{r}{6.6cm}
\centerline{\includegraphics[width=6 cm,height=3 cm]{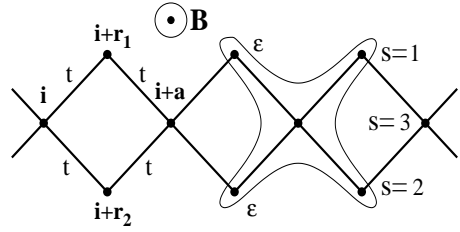}}
\caption{Diamond Hubbard chain for $t_{\perp}=t_{\parallel}=0$. The
thin line depicts the localized eigenstate corresponding to $\hat
C^{\dagger}_{3, {\bf i},\sigma}$.} \label{fig:3}
\end{wrapfigure}
%%%%%%%%%%%%%%%%%%%%%%%%%%%%%%%%%%%%%%%%%%%%%%%%%%%%%%%%%%%%%%%%%%%%%%%%%

We first consider the parameters
$t_{\perp}=t_{\parallel}=0,\delta=\frac{\pi}{2}$, in which case the
single-electron band structure consists of three flat bands (see
Fig. \ref{fig:2}a). This corresponds to the ''Aharonov-Bohm cage'' limit
analyzed by Vidal {\it et al.} \cite{Dia1} for the two electron
problem with $\varepsilon=0$ (see also Fig. \ref{fig:3}). Based on their
finding that excited singlet eigenstates are localized for $U=0$ but
delocalized for $U > 0$, Vidal {\it et al.} conjectured that such
delocalized states should emerge also at finite electron densities.
Below we deduce indeed exact ground states for the diamond Hubbard
chain at finite densities.

Following the strategy outlined in \S 2 one first transforms the
kinetic part of the Hamiltonian into positive semidefinite form by
introducing new canonical fermionic operators $\hat C_{\nu,{\bf
i},\sigma}$ in real space (for details see Ref.\cite{re15}); these
operators represent localized Wannier eigenstates (see Fig.
\ref{fig:3}). With the energies $E_2=\varepsilon$, $E_{2\pm
1}=(\varepsilon \mp \sqrt{{\varepsilon}^2 +4})/2$ one obtains
\begin{equation}
{\hat H}_0=\sum_{\sigma,{\bf i}=1}^{N_c}\sum_{\nu=1}^3 E_\nu
\hat C^{\dagger}_{\nu,{\bf i},\sigma}\hat C_{\nu,{\bf i},\sigma}\, .
\label{dia4}
\end{equation}
Since $\hat C^{\dagger}_{\nu,{\bf i},\sigma}\hat C_{\nu,{\bf
i},\sigma}$ and $\hat H_U$  are positive semidefinite operators the
ground state of $\hat H$ for $N_c \geq N$, $U > 0$ becomes
\begin{eqnarray}
|\Psi^{I}_g(N)\rangle = \prod_{{\bf i}=1}^N \hat C^{\dagger}_{3,{\bf i},
\sigma_{\bf i}}|0\rangle, \quad  E^{I}_g = E_3 N\, .
\label{dia5}
\end{eqnarray}

\begin{wrapfigure}{l}{6.6cm}
\centerline{\includegraphics[height=6.cm]{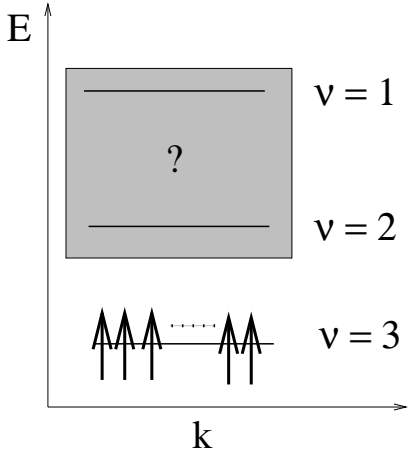}}
\caption{Ferromagnetic flat-band solution on the diamond Hubbard
chain obtained for $t_{\perp}=t_{\parallel}=0$; the ground-state
result does not provide information about higher lying states as
indicated by the question mark.} \label{fig:4}
\end{wrapfigure}

For $n=1/3$ one obtains a fully saturated ferromagnetic ground state
in the lowest band (see Fig. \ref{fig:4}) which represents an
explicit realization of Mielke-Tasaki's flat band ferromagnetism
\cite{Dia2}. The strategy for the proof of the uniqueness of this
ground state is discussed in Appendix A. For $n<1/3$ only the
Wannier states with a spatial overlap have the same spin, the
(highly degenerate) ground state hence consists of ferromagnetic
clusters of arbitrary spin orientation. Since $\hat H_U
|\Psi^{I}_g(N)\rangle =0$ and the kinetic part of $\hat H$ is
diagonal in real space, the ground state (\ref{dia5}) is localized.
The explicit proof of the uniqueness of (\ref{dia5}) for $n<1/3$ is
presented in Appendix B.

The flat-band ferromagnetism found here realizes early ideas of
Gutzwiller\cite{Dia3} and Kanamori\cite{Dia4} from 1963  about the
origin of ferromagnetism. They argued that the ferromagnetic
orientation of the electronic spins is due to the fact that the
local interaction is thereby completely suppressed thus lowering the
energy greatly.
%%%%%%%%%%%%%%%%%%%%%%%%%%%%%%%%%%%%%%%%%%%%%%%%%%%%%%%%%%%%%%%%%%%%%%%%
% FIGURE 4
%%%%%%%%%%%%%%%%%%%%%%%%%%%%%%%%%%%%%%%%%%%%%%%%%%%%%%%%%%%%%%%%%%%%%%%%
%%%%%%%%%%%%%%%%%%%%%%%%%%%%%%%%%%%%%%%%%%%%%%%%%%%%%%%%%%%%%%%%%%%%%
However, they were not able to compute the associated loss in
kinetic energy of this highly correlated state. In the present case
of flat-band ferromagnetism the latter effect poses no problem since
the kinetic energy is zero anyway.

We note that at $U
> 0$ the lowest band is flat only for $\delta=\pi/2$, but is
dispersive for $\delta=\pi$, or $\delta =0$. In the latter situation
the system is most probably conducting for incommensurate fillings.
Hence one encounters here a magnetic field induced metal-insulator
transition.

\subsection{Correlated half-metal}
The previous results suggest that itinerant states are easier to
realize at $\delta \ne\pi/2$. To analyze this problem we investigate
below the special case $t_{\parallel} >0$, $\epsilon=-t_{\perp}
+t^{-1}_{\perp}/2$, and $\delta=\pi$; the single-electron band
structure has a dispersive lowest band and two upper flat bands (see
Fig. \ref{fig:2}b).

\subsubsection{Transformation of ${\hat H}$ into positive semidefinite form}
For the rewriting of $\hat H_0$ one defines first the non-canonical fermionic
operators
\begin{eqnarray}
\hat A_{{\bf i},\sigma}=a_1 \hat c_{{\bf i},\sigma} + a_2
\hat c_{{\bf i}+{\bf r}_2,\sigma} + a_3 \hat c_{{\bf i}+{\bf a},\sigma} +
a_4 \hat c_{{\bf i}+{\bf r}_1,\sigma}\, ,
\label{dia6}
\end{eqnarray}
which fulfill $(\hat A_{{\bf i},\sigma})^2=0$ and $\{\hat A_{{\bf i},\sigma},
\hat A^{\dagger}_{{\bf j},\sigma} \} \ne \delta_{{\bf i},{\bf j}}$.
Based on (\ref{dia6}) one obtains
\begin{eqnarray}
\hat A^{\dagger}_{{\bf i},\sigma} \hat A_{{\bf i},\sigma} &=& (
a^{*}_2 a_1 \hat c^{\dagger}_{{\bf i}+{\bf r}_2,\sigma}
\hat c_{{\bf i},\sigma} + a^{*}_3 a_2 \hat c^{\dagger}_{{\bf i}+{\bf a},
\sigma} \hat c_{{\bf i}+{\bf r}_2,\sigma} + a^{*}_4 a_3 \hat c^{\dagger}_{
{\bf i}+{\bf r}_1,\sigma} \hat c_{{\bf i}+{\bf a},\sigma} +
a^{*}_1 a_4 \hat c^{\dagger}_{{\bf i},\sigma} \hat c_{{\bf i}+{\bf r}_1,
\sigma}
\nonumber\\
&+& a^{*}_2 a_4 \hat c^{\dagger}_{{\bf i}+{\bf r}_2,\sigma}
\hat c_{{\bf i}+{\bf r}_1,\sigma} + a^{*}_3 a_1 \hat c^{\dagger}_{{\bf i}+
{\bf a},\sigma} \hat c_{{\bf i},\sigma} + H.c. )
\nonumber\\
&+&
|a_1|^2 \hat n_{{\bf i},\sigma} + |a_2|^2 \hat n_{{\bf i}+{\bf r}_2,\sigma} +
|a_3|^2 \hat n_{{\bf i}+{\bf a},\sigma} + |a_4|^2
\hat n_{{\bf i}+{\bf r}_1,\sigma} .
\label{dia7}
\end{eqnarray}
As discussed in Sec. 2.2. the coefficients $a_1$, $a_2$, $a_3$,
$a_4$ can be determined in different ways, namely by either
requiring  $-\sum_{{\bf i},\sigma}\hat A^{\dagger}_{{\bf i},\sigma}
\hat A_{{\bf i},\sigma}=\hat H_0 $ or $+\sum_{{\bf i},\sigma}\hat
A^{\dagger}_{{\bf i},\sigma} \hat A_{{\bf i},\sigma}=\hat H_0 $.
Here we choose the former condition which will enable us to
construct exact ground states for $N \geq 4N_{c}$ (see Sec. 3.2.2.).
With the requirement $-\sum_{{\bf i},\sigma}\hat A^{\dagger}_{{\bf
i},\sigma} \hat A_{{\bf i},\sigma}=\hat H_0 $ one arrives at the
matching conditions
\begin{eqnarray}
&&a^{*}_2 a_1 = a^{*}_3 a_2 = a^{*}_4 a_3 = a^{*}_1 a_4 = -t e^{i \frac{
\delta}{2}}, \quad
a^{*}_2 a_4 = - t_{\perp},
\quad
a^{*}_3 a_1 = - t_{\parallel},
\nonumber\\
&&|a_{1}|^2 + |a_{3}|^2 = \epsilon + |a_{2}|^2 = \epsilon + |a_{4}|^2,
\label{dia9}
\end{eqnarray}
from which
\begin{equation}
\hat A_{{\bf i},\sigma}=\sqrt{t_{\parallel}}[\hat c_{{\bf
i},\sigma}-\hat c_{{\bf i}+{\bf a},\sigma}- 2 t_{\perp}
e^{i\frac{\delta}{2}} (\hat c_{{\bf i}+{\bf r}_1, \sigma} - \hat
c_{{\bf i}+{\bf r}_2,\sigma})] \, \label{dia10}
\end{equation}
follows. Based on (\ref{dia7} -\ref{dia10}), ${\hat H}_0$ is rewritten as
\begin{eqnarray}
\hat H_0 = - \sum_{{\bf i},\sigma} \hat A^{\dagger}_{{\bf i},\sigma}
\hat A_{{\bf i},\sigma} = + \sum_{{\bf i},\sigma} \hat A_{{\bf i},\sigma}
\hat A^{\dagger}_{{\bf i},\sigma} - 2 N_c \sum_{m=1}^4 |a_m|^2 .
\label{dia11}
\end{eqnarray}
For the interaction part $\hat H_U$, one uses
\begin{eqnarray}
\hat H_U = U \sum_{\bf i}^{3 N_c} \hat n_{{\bf i},\uparrow}
\hat n_{{\bf i},\downarrow} = U \hat P + U \hat N - U N_c, \quad
\hat P = \sum_{\bf i} \hat P_{\bf i}\, .
\label{dia12}
\end{eqnarray}
${\hat N}$ is the particle number operator, and $\hat P_{\bf i}=(\hat
n_{{\bf i},\uparrow}-1)(\hat n_{{\bf i},\downarrow}-1)$
is a positive semidefinite operator, which gives zero for at least one
electron on site ${\bf i}$ and 1 for an unoccupied site ${\bf i}$.
Consequently, $\hat H$ has been transformed into the positive semidefinite
form
\begin{eqnarray}
\hat H = \sum_{{\bf i},\sigma} \hat A_{{\bf i},\sigma}
\hat A^{\dagger}_{{\bf i},\sigma} + U \hat P + E^{II}_g ,
\label{dia13}
\end{eqnarray}
where
$E^{II}_g =(\epsilon + U + t_{\perp}) N-N_c (3U+4 t_{\perp} + 1/t_{\perp})$.

\subsubsection{Construction of the ground state}
As discussed in Sec. 2.2, the ground state $|\Psi_g\rangle$ for $U >
0$ is deduced such to satisfy $\hat A^{\dagger}_{{\bf
i},\sigma}|\Psi_g\rangle =0$ and $\hat P |\Psi_g \rangle =0$, from
where indeed
\begin{eqnarray}
\hat H |\Psi_g \rangle = E_g |\Psi_g \rangle
\label{dia15}
\end{eqnarray}
follows. Since $(\hat A^{\dagger}_{{\bf i},\sigma})^2 =0$ one finds
\begin{eqnarray}
|\Psi^{II}_g (4N_c)\rangle \propto \prod_{\bf i} \hat A^{\dagger}_{{\bf i},
-\sigma} \hat A^{\dagger}_{{\bf i},\sigma} |0\rangle .
\label{dia16}
\end{eqnarray}
In order to satisfy $\hat P |\Psi_g \rangle =0$ as well, which requires one
electron on each site, besides the contribution from (\ref{dia16}) one has
to introduce in the ground state wavefunction also the operator
\begin{eqnarray}
\hat F^{\dagger}_{\sigma} = \prod_{\bf i} [\hat c^{\dagger}_{{\bf i}+{\bf r}_{
s_{{\bf i},1}},\sigma} \hat c^{\dagger}_{{\bf i}+{\bf r}_{s_{{\bf i},2}},
\sigma}],
\label{dia17}
\end{eqnarray}
which creates two electrons with fixed spin $\sigma$ on arbitrary sites of
each unit cell. Hence, the unnormalized ground state becomes
\begin{eqnarray}
|\Psi^{II}_g (4N_c)\rangle = c [\prod_{\bf i} \hat A^{\dagger}_{{\bf i},
-\sigma} \hat A^{\dagger}_{{\bf i},\sigma}] \hat F^{\dagger}_{\sigma}
|0\rangle ,
\label{dia18}
\end{eqnarray}
where $c$ is a normalization constant. The contribution in
(\ref{dia16}) creates one $\sigma$ electron, and an electron with
spin $-\sigma$ in each unit cell, while $\hat F^{\dagger}_{\sigma}$
creates two electrons with spin $\sigma$ on arbitrary sites of each
unit cell. Consequently the ground state in (\ref{dia18})
corresponds to $N=4N_c$, $n=4/3$, $n_{\sigma}=N_\sigma /3N_c=1$,
$n_{-\sigma}=1/3$. The proof of the uniqueness of (\ref{dia18}) is
presented in detail in Appendix A. In conclusion, (\ref{dia18}) is
the unique ground state.

Concerning the physical properties of the ground state one observes
that in (\ref{dia18}) each lattice site is occupied by one
$\sigma$-electron; hence, the $\sigma$-electrons are localized. The
$-\sigma$ electrons are spatially extended, but localized for $N_c
\to\infty$. This is inferred from the ground-state expectation value
of the hopping term $\Gamma_{{\bf r},-\sigma}=\langle \hat
c^{\dagger}_{{\bf j},-\sigma}\hat c^{\dagger}_{{\bf j}+{\bf
r},-\sigma}+ H.c. \rangle$, which in the limit $N_c \to\infty$, for
$r/a=m$, is calculated as
\begin{eqnarray}
\Gamma_{m,-\sigma}= \frac{(-1)^m}{\sqrt{1+1/t_{\perp}}} e^{-\frac{m}{
\xi_{-\sigma}}}\, .
\label{dia19}
\end{eqnarray}
The one-particle localization length $\xi_{-\sigma}$ in (\ref{dia19}) increases
almost linearly with $1/t_{\perp}$ \cite{re15}.

\subsubsection{Solution for $n> 4/3$}
For electron densities $n > 4/3$ there are $N > 4N_c$ electrons. No
$\sigma$-electrons but $\Delta N$ $-\sigma$-electrons can be added
to the system such that $n_{\sigma}=1$, $n_{-\sigma}=1/3 + \Delta
N/N_c$. The ground state then becomes
\begin{eqnarray}
|\Psi^{II}_g (4N_c + \Delta N)\rangle = \prod_{\alpha=1}^{\Delta N}
\hat c^{\dagger}_{n_{\alpha},{\bf k}_{\alpha},-\sigma}
|\Psi_g^{II}(4N_c) \rangle,
\label{dia20}
\end{eqnarray}
where $n_{\alpha}$ can be arbitrarily chosen from $s=1,2,3$. Eq.
(\ref{dia20}) contains plane wave-type contributions from the
$-\sigma$-electrons. Consequently, the additional $\Delta N$
$-\sigma$ electrons are itinerant. Indeed, this follows explicit
from the expectation value of the hopping term $\Gamma_{\bf
r,-\sigma}$. For $|{\bf r}/{\bf a}|\gg 1$, $\Delta N=1$, $s=1$, one
finds $\Gamma_{{\bf r},-\sigma}/\Gamma^{(1)}_{\bf
r,-\sigma}=1-[2+(1- \cos ak)/t_{\perp}]^{-1}$. Here
$\Gamma^{(1)}_{\bf r,-\sigma}$ is the plane wave result and $k$ is
the norm of the momentum of the electron added above $n=4/3$. We
note that the momenta ${\bf k}_{\alpha}$ in (\ref{dia20}) are
arbitrary. Hence it is not possible to define a Fermi surface, and
the ground state therefore corresponds to a non-Fermi liquid.

In the density range $4/3 < n < 5/3$ the ground state has $3N_c$
immobile $\sigma$ electrons and $N_c$ electrons with spin $-\sigma$
confined to their localized Wannier functions. For this reason the
ground state has a high geometric degeneracy; its uniqueness is
discussed in Appendix B.

Only the $\Delta N$ conducting $-\sigma$ electrons are itinerant,
leading to metallic behavior with a low carrier density and a low
spin polarization. Since the conduction through this correlated
half-metal involves only electrons of one spin species, such a
system may serve as a spin-valve device. Since $U > 0$ is required
for this property one encounters here a correlation induced
localization-delocalization transition to a half-metal. The
magnetization behaves as $M\propto (1-\Delta N/N_c) \to 0$ for
$\Delta N \to N_c$. in a finite magnetic field. The transition can
also be induced by tuning the local potential $\varepsilon$. At
$n=5/3$ the ground state becomes nonmagnetic and localized in the
thermodynamic limit, while at $n > 5/3$ one finds a nonmagnetic but
conducting ground state.

\subsection{Exact ground states for general magnetic flux}
It is also possible to construct exact ground states for general
magnetic flux values including also the ${\bf B}={\bf 0}$ case, for
example at $\delta\in (-\pi/2, \pi/2)$ with $t_{\parallel}=0$,
$t_{\perp} < 0$, $b\equiv -\cos \delta/t_{\perp}$, $\varepsilon =
b-b^{-1}$, and $n \geq 5/3$ \cite{re15}; the single-electron band
structure has two dispersive bands and an upper flat band (see Fig.
\ref{fig:2}c). For these hopping and potential parameters one obtains a
localized nonmagnetic ground state for $\delta =0$ over an extended
range of densities $n \geq 5/3$. By contrast, at finite magnetic
fields the ground state is a nonsaturated ferromagnet. This
localized state at $n=5/3$ is insulating, but gapless and itinerant
for $n > 5/3$. For the latter densities the majority spin
$\sigma$-electrons are immobile and only the $-\sigma$ electrons are
itinerant. Therefore, by varying the external magnetic field and the
sublattice potential one can tune from a localized, nonmagnetic
ground state to a nonsaturated ferromagnet in the density range $n
\geq 5/3$; the ferromagnetic state is insulating at $n=5/3$, but
conducting for $n > 5/3$.

We finally note the properties derived above do not depend on the Zeeman
coupling. Only the solution in Sec. 3.1 for $n <1/3$ is altered by the
Zeeman term, whose presence leads to a fully spin aligned ferromagnetic
solution.

\section{Exact many-electron ground states on triangle Hubbard chains}

\begin{wrapfigure}{r}{6.6cm}
\centerline{\includegraphics[width=6 cm,height=2.5
cm]{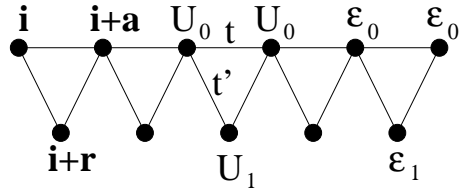}} \caption{Triangle Hubbard chain} \label{fig:5}
\end{wrapfigure}

The procedure for the construction of exact ground states on the diamond
Hubbard chain is now applied to a Hubbard chain composed of corner-sharing
triangles (see Fig. \ref{fig:5}). Triangles are well-known to frustrate
non-ferromagnetic magnetic order. Hence interacting electrons on
lattices with triangles as structural sub-units are expected to favor
ferromagnetic ground states.

%%%%%%%%%%%%%%%%%%%%%%%%%%%%%%%%%%%%%%%%%%%%%%%%%%%%%%%%%%%%%%%%%%%%%%%
% FIGURE 5
%%%%%%%%%%%%%%%%%%%%%%%%%%%%%%%%%%%%%%%%%%%%%%%%%%%%%%%%%%%%%%%%%%%%%%%
%%%%%%%%%%%%%%%%%%%%%%%%%%%%%%%%%%%%%%%%%%%%%%%%%%%%%%%%%%%%%%%%%%%%%%%
The triangle Hubbard chain in Fig. 5 has two sites per cell which
leads to two energy bands. The number of electrons, cells, and
electron density are here denoted by $N$, $N_c$ and $n=N/(2N_c)$,
respectively. This model was previously investigated in detail by
Penc {\it et al.} \cite{Dia5}, Fazekas {\it et al.} \cite{Dia6},
M\"uller-Hartmann \cite{Dia7} and Derzhko {\it et al.} \cite{Dia8}
with a variety of analytical and numerical techniques.

The Hamiltonian of the triangle chain $\hat H^{tri} =\hat
H_{0}^{tri} + \hat H_{U}^{tri}$ has the form
\begin{eqnarray}
&&\hat H_{0}^{tri} = \sum_{\sigma} \sum_{{\bf i}=1}^{N_c} \: [ \: (t
\hat c^{\dagger}_{ {\bf i},\sigma} \hat c_{{\bf i}+{\bf a},\sigma} +
t' (\hat c^{\dagger}_{{\bf i},\sigma} \hat c_{{\bf i}+{\bf r},\sigma} +
\hat c^{\dagger}_{{\bf i}+{\bf a},\sigma} \hat c_{{\bf i}+{\bf r},\sigma}) +
H.c.) \nonumber \\
&&\hskip2.8cm + \varepsilon_0 \hat c^{\dagger}_{{\bf i},\sigma} \hat c_{{\bf
i},\sigma}
+ \varepsilon_1
\hat c^{\dagger}_{{\bf i}+{\bf r},\sigma} \hat c_{{\bf i}+{\bf r},\sigma}
\: \: ],
\nonumber\\
&&\hat H_{U}^{tri} = \sum_{{\bf i}} [ U_0 \hat n_{{\bf i},\uparrow}
\hat n_{{\bf i},\downarrow} + U_1 \hat n_{{\bf i}+{\bf r},\uparrow}
\hat n_{{\bf i}+{\bf r},\downarrow} ].
\label{eqtri1}
\end{eqnarray}
The model parameters are explained in Fig. \ref{fig:5}. In the following we
will construct exact ground states in the subspace of parameters
defined by the relation $(t')^2/t=\varepsilon_1 - \varepsilon_0 + 2t$,
$\varepsilon_1 - \varepsilon_0 > -2t$, $t > 0$, for which the lowest
bare band is flat.

\subsection{Ground states for the triangle Hubbard chain}

\subsubsection{Ground state for $U_0$, $U_1 > 0$}

%%%%%%%%%%%%%%%%%%%%%%%%%%%%%%%%%%%%%%%%%%%%%%%%%%%%%%%%%%%%%%%%%%%%%%%
% FIGURE 6a,b
%%%%%%%%%%%%%%%%%%%%%%%%%%%%%%%%%%%%%%%%%%%%%%%%%%%%%%%%%%%%%%%%%%%%%%%
\begin{figure}
\centerline{\includegraphics[width=12 cm,height=6
cm]{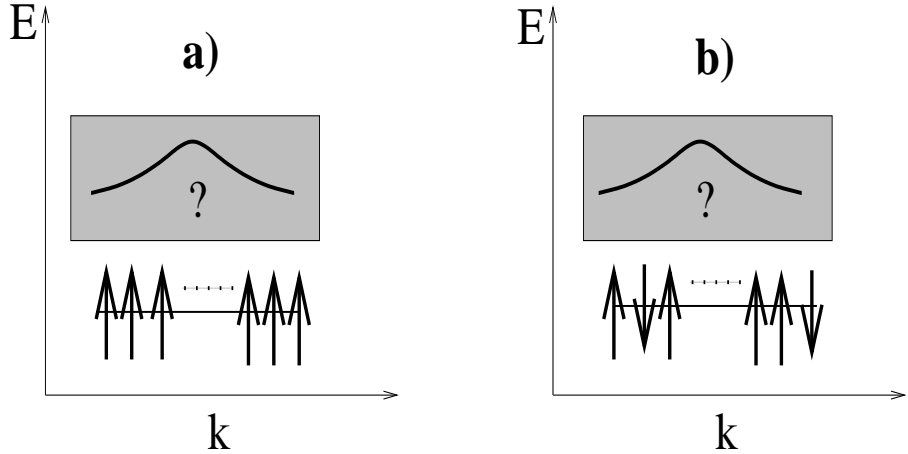}} \caption{Ground-state configuration on the
triangle Hubbard chain for (a) $U_0,U_1>0$, and (b) $U_0>0, U_1=0$.
The ground-state results do not provide information about higher
lying states as indicated by the question marks.} \label{fig:6}
\end{figure}
%%%%%%%%%%%%%%%%%%%%%%%%%%%%%%%%%%%%%%%%%%%%%%%%%%%%%%%%%%%%%%%%%%%%%%%%

Employing the operators
\begin{eqnarray}
\hat A_{{\bf i},\sigma}= \sqrt{t} [ \: \hat c_{{\bf i},\sigma} +
\hat c_{{\bf i}+{\bf a},\sigma} + (t'/t) \hat c_{{\bf i}+{\bf
r},\sigma} ]
\end{eqnarray}
the Hamiltonian (\ref{eqtri1}) can be transformed into positive
semidefinite form:
\begin{eqnarray}
\hat H^{tri} = \sum_{{\bf i},\sigma} \hat A^{\dagger}_{{\bf
i},\sigma} \hat A_{{\bf i},\sigma} +  K \hat N + H_U^{tri}.
\label{eq1.9}
\end{eqnarray}
Here $\hat N$ is the particle number operator and $K=\varepsilon_0-2t$.
To construct the ground state one has to find operators
$\hat B^{\dagger}_{{\bf i},\sigma}$ which satisfy $\{ \hat A_{{\bf
i},\sigma} , \hat B^{\dagger}_{{\bf i}',\sigma'} \} = 0$ for all
values of the indices. The ground state then has the form
\begin{eqnarray}
|\Psi^{tri}(N)\rangle = \prod_{\bf i}^N \hat B^{\dagger}_{{\bf
i},\sigma_{\bf i}} |0\rangle,\label{tri_GS}
\end{eqnarray}
with $\hat B^{\dagger}_{{\bf i},\sigma}= [ \hat c^{\dagger}_{{\bf
i}- {\bf a}+{\bf r},\sigma} + \hat c^{\dagger}_{{\bf i}+{\bf
r},\sigma} - (t'/t) \hat c^{\dagger}_{{\bf i},\sigma} ]$, $N \leq
N_c$. For $n < 1/2$ there exist ferromagnetic clusters, and only
operators $\hat B^{\dagger}_{{\bf i},\sigma_{\bf i}}$ which touch
have the same spin index. For $n=1/2$ all clusters touch such that a
fully saturated ferromagnetic state is obtained as shown in Fig. \ref{fig:6}a
for the $U_1 > 0$ case. The ground state
corresponds to a particular realization of flat-band ferromagnetism
\cite{Dia2}, for which Derzhko {\it et al.} \cite{Dia8} calculated
the thermodynamic properties.

\subsubsection{Ground state for $U_0 > 0$, $U_1 = 0$}

The  ground state has the same form as (\ref{tri_GS}), but now it is
\emph{non}-magnetic even for $n=1/2$ (see Fig. \ref{fig:6}b).
This is due to the fact that,
although the connectivity condition is fulfilled (i.e., $\hat
B^{\dagger}_{{\bf i},\sigma}$ and $\hat B^{\dagger}_{{\bf i}+ {\bf
a},\sigma}$ overlap at site ${\bf i}+{\bf r}$) there is no
interaction at the connectivity point ($U_1=0$).
The fulfillment of the connectivity condition for the Wannier
functions is therefore not sufficient for the ground state to be
ferromagnetic. In addition a finite Hubbard repulsion is needed
\emph{at} the connectivity points to guarantee a correlation between
the electrons.

\subsubsection{Ground state for $U_0 = 0, U_1 > 0$}

The ground state has again the form (\ref{tri_GS}), but now it
is a fully saturated ferromagnetic state at $n=1/2$ even for
$U_0=0$, since the Hubbard interaction is non-zero at the
connectivity points ($U_1>0$). We thus see that for flat-band
ferromagnetism to occur the Hubbard interaction is needed
\emph{only} at the connectivity points. The ferromagnetic states
discussed in Sec. 4.1.1 and Sec. 4.1.3. are localized in the
thermodynamic limit.

\subsection{Mapping to the one-dimensional periodic Anderson model (PAM)}

By the following change of notation in (\ref{eqtri1})
\begin{eqnarray}
\hat c_{{\bf i}\sigma} \rightarrow \hat d_{{\bf i},\sigma}, \hskip0.2cm
\hat c_{{\bf i}+{\bf r},\sigma} \rightarrow \hat f_{{\bf i},\sigma},\hskip0.2cm
t' \rightarrow V, \hskip0.2cm\varepsilon_1 \rightarrow E_f, \hskip0.2cm
U_1 \rightarrow U,
\end{eqnarray}
and setting $\varepsilon_0=U_0=0$ the triangle Hubbard chain transforms
into a one-dimensional PAM with zero on-site hybridization and
nearest-neighbor hybridization $V$:
\begin{eqnarray}
\hat H^{PAM}&=& \sum_{\sigma} \sum_{{\bf i}=1}^{N_c} (t \hat
d^{\dagger}_{ {\bf i},\sigma} \hat d_{{\bf i}+{\bf a},\sigma} + H.c
) + E_f \hat f^{\dagger}_{{\bf i}+{\bf r},\sigma} \hat f_{{\bf
i}+{\bf r},\sigma}
\nonumber\\
&+&V(\hat d^{\dagger}_{{\bf i},\sigma} \hat f_{{\bf i}+{\bf r},\sigma}
+ \hat d^{\dagger}_{{\bf i}+{\bf a},\sigma} \hat f_{{\bf i}+{\bf
r},\sigma} + H.c.)] + U \sum_{{\bf i}} \hat n^f_{{\bf i}+{\bf
r},\uparrow} \hat n^f_{{\bf i}+{\bf r},\downarrow}.
\label{triangleeq1}
\end{eqnarray}
 A lowest bare flat band is obtained for $V^2/t=E_f +2t$, $E_f > -2t$, $t > 0$.

\subsubsection{Exact ground state for the one-dimensional PAM}

In the parameter regime $V^2/t=E_f +2t$, $E_f > -2t$, $t > 0$ the exact
ground state for the one-dimensional PAM is obtained from
(\ref{tri_GS}) as
\begin{eqnarray}
|\Psi_g^{PAM}(N=N_c)\rangle = \prod_{{\bf i}=1}^{N_c} [ \hat
f^{\dagger}_{{\bf i} -{\bf a}+{\bf r},\sigma} + \hat
f^{\dagger}_{{\bf i}+{\bf r},\sigma}- \frac{V}{t} \hat
d^{\dagger}_{{\bf i},\sigma} ] |0 \rangle. \label{tri3}
\end{eqnarray}

\subsubsection{Itinerant ferromagnetism of finite chains}

\begin{wrapfigure}{r}{6.60cm}
\centerline{\includegraphics[width=6.6cm]{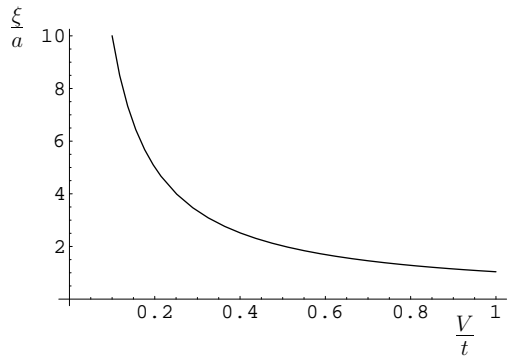}}
\caption{Localization length $\xi$ of the $d$-electrons of the
one-dimensional PAM, (\ref{triangleeq1}), as a function of the
nearest-neighbor hybridization strength $V$.} \label{fig:7}
\end{wrapfigure}

In analogy with (\ref{dia19}) the localization length $\xi$ of the
$d$-electrons can be calculated from the expression for the
expectation value of the hopping term $\Gamma_{{\bf r}}$.  Its
dependence on the nearest-neighbor hybridization strength $V$ is
given by
\begin{eqnarray}
\xi =\frac{a}{\ln [1 + \frac{V^2}{2t^2} (1-\sqrt{1+
\frac{4t^2}{V^2}} ) ]}, \hskip0.5cm
\label{tri4}
\end{eqnarray}
as shown in Fig. \ref{fig:7}. Here $a$ is the distance between sites
at the base of the triangle chain (see Fig. \ref{fig:5}). Although
flat-band ferromagnetism in the thermodynamic limit usually
corresponds to localized electrons, the localization length can be
so large that it exceeds the length of a finite sample. In realistic
situations the exact ground state of the finite chain therefore
shows \emph{itinerant ferromagnetism}.

\subsection{Application to $f$-electron ferromagnetism in CeRh$_3$B$_2$}

As pointed out by Kono and Kuramoto \cite{Dia10} the one-dimensional
PAM, (\ref{triangleeq1}), may be employed to model the $4f$-electron compound
CeRh$_3$B$_2$. This material is ferromagnetic below $T_c=120$ K. Its
properties are interesting for several reasons: (i) the
ferromagnetic state cannot be explained by a RKKY interaction, (ii)
the $4f$ moment has the remarkably small value  of 0.45 $\mu_B$
compared to the free Ce$^{3+}$ moment of 2.14 $\mu_B$, and (iii)
the Curie temperature $T_c$ has the highest value among all known Ce
compounds with non-magnetic elements.

%%%%%%%%%%%%%%%%%%%%%%%%%%%%%%%%%%%%%%%%%%%%%%%%%%%%%%%%%%%%%%%%%%%%%%%%%%
% FIGURE 7.
%%%%%%%%%%%%%%%%%%%%%%%%%%%%%%%%%%%%%%%%%%%%%%%%%%%%%%%%%%%%%%%%%%%%%%%%%%
%%%%%%%%%%%%%%%%%%%%%%%%%%%%%%%%%%%%%%%%%%%%%%%%%%%%%%%%%%%%%%%%%%%%%%%%%%

The lattice structure of the class of rare-earth (RE) compounds
RERh$_3$B$_2$ was determined by Yamada {\it et al.} \cite{Dia9}. A
cut through the (a,c)-plane of the Cerium system (RE = Ce) is shown
in Fig. 8. The remarkable difference between the distances of the Ce
atoms along the a- and the c-direction, respectively, makes this
material a quasi one-dimensional two-band system \cite{1D}. For this
reason Kono and Kuramoto \cite{Dia10} modeled CeRh$_3$B$_2$ by the
triangle PAM chain shown in Fig. 9. The parameters entering the
model are $t=0.34$ eV, $V=0.24$ eV, $E_f = -0.714$ eV, $U = 7$ eV,
$n=0.55$, as obtained from band-structure calculations and X-ray
absorption spectroscopy (we note that the density $n=0.55$
corresponds to twice the value, i.e., $n=1.1$, in the notation of
Ref. \cite{Dia10}).

%%%%%%%%%%%%%%%%%%%%%%%%%%%%%%%%%%%%%%%%%%%%%%%%%%%%%%%%%%%%%%%%%%%%%%%
%% FIGURES 8-9
%%%%%%%%%%%%%%%%%%%%%%%%%%%%%%%%%%%%%%%%%%%%%%%%%%%%%%%%%%%%%%%%%%%%%%%
\begin{figure}
\parbox{\halftext}{
\centerline{\includegraphics[width=6 cm,height=6 cm]{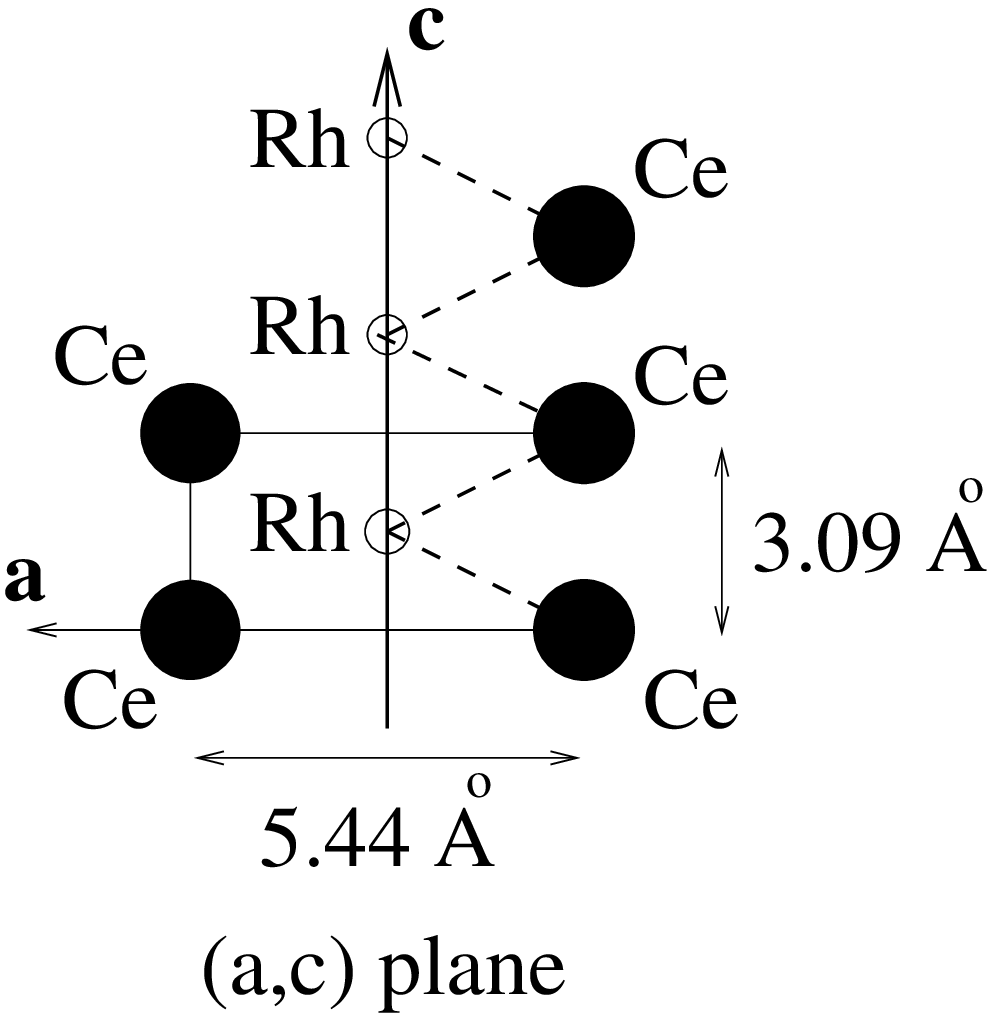}}
\caption{Cut through the (a,c)-plane of the lattice structure of
CeRh$_3$B$_2$ based on the parameters determined in Ref.
\cite{Dia9}.}} \label{fig:8} \hfill
\parbox{\halftext}{
\centerline{\includegraphics[width=6.4cm]{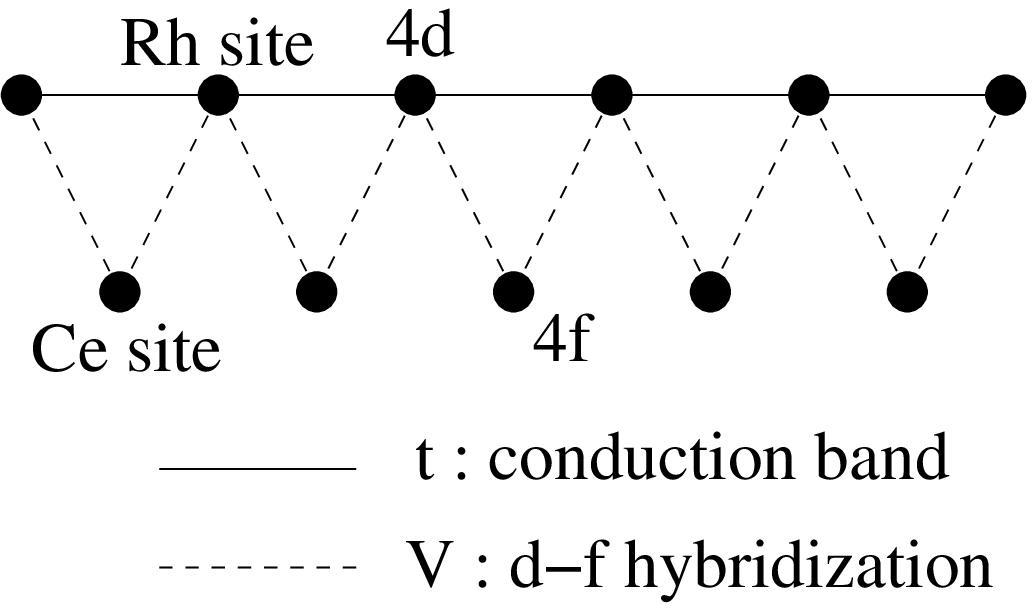}}
\caption{One-dimensional PAM
used to model CeRh$_3$B$_2$ \cite{Dia10}.}}
\label{fig:9}
\end{figure}
%%%%%%%%%%%%%%%%%%%%%%%%%%%%%%%%%%%%%%%%%%%%%%%%%%%%%%%%%%%%%%%%%%%%%%%

\subsubsection{Variational results for the one-dimensional PAM}

Kono and Kuramoto \cite{Dia10} computed the band structure of
CeRh$_3$B$_2$ by constructing a Gutzwiller variational wave function
and using variational Monte Carlo to determine the ground state.
They found the lowest band of CeRh$_3$B$_2$ to be almost perfectly
flat and fully polarized, and the second band dispersive and
almost unpolarized. It should be stressed that the shape of the
lowest quasiparticle band of interacting electrons depends on $U$.

\subsubsection{Exact ground state of the one-dimensional PAM}

We can use the exact ground state of the one-dimensional PAM,
(\ref{tri3}), to make direct contact with the $f$-electron
ferromagnetism found in CeRh$_3$B$_2$. Namely, this ground state
represents a saturated flat-band ferromagnetic state of the
$f$-electrons for electron density $n=1/2$ and arbitrary $U>0$,
provided the condition $V^2/t=E_f +2t$, $E_f > -2t$, $t>0$ holds.
The condition is indeed fulfilled for the parameter values $t=0.34$
eV, $V=0.23$ eV, $E_f= -0.52$ eV, which are very close to the
experimentally determined values. In this case the lowest band is
perfectly flat for all $U>0$. Apparently the ferromagnetism of the
$f$-electrons in CeRh$_3$B$_2$ is due to the presence of an (almost)
flat lowest band.

\subsubsection{Magnetic moments}

The experimentally observed magnetic moment of the $f$-electrons is
$m_f = 0.45$ $\mu_B$ \cite{Dia11}. The variational result
\cite{Dia10} yields $m_f = 0.94$ $\mu_B$, while the exact ground
state discussed above leads to a value of $m_f = 0.68$ $\mu_B$ which
is even closer to the experimental value. Both theoretical values
are somewhat higher than the experimental $f$ moment. This may be
due to the fact that the theoretical approaches do not include the
spin-orbit interaction which causes an antiferromagnetic coupling
between the $d$- and the $f$-electrons.

\section{Conclusions}

Despite its simplicity, the diamond Hubbard chain displays
remarkably complex physical properties with insulating, conducting,
and fully or partially spin polarized states. The constructed exact
many-electron ground states revealed flat-band ferromagnetism,
correlated half-metallic behavior, and metal-insulator transitions
in the diamond Hubbard chain. The virtue of tuning fundamentally
different ground states through external magnetic fields, site
selective potentials, or the electron density, points to new
prospects for designing electronic devices, which allow to switch
between different ground states, including states that operate as a
spin valve. We applied the general strategy for the construction of
exact ground states also to triangle Hubbard chains. The mapping of
the triangle Hubbard chain to the one-dimensional periodic Anderson
model within a restricted parameter set provided a theory for
$f$-electron ferromagnetism in CeRh$_3$B$_2$, which is based on the
physical properties of an exact ground state.

Further applications of the construction of exact ground states by
casting the Hamiltonian into positive semidefinite form are possible
for model systems with selected geometries in artificially designed
structures. In particular, we constructed exact ground states for
pentagon Hubbard chains. These results will be presented elsewhere.

\section*{Acknowledgements}

We thank F. Mila, Y. Kuramoto, and F. C. Zhang for discussions regarding the
uniqueness of the solutions. Support by the Hungarian Research Fund through
Contracts No. OTKA-T48782, 76821, the Alexander von Humboldt Foundation, and
the Deutsche Forschungsgemeinschaft through SFB 484 is gratefully acknowledged.

%%%%%%%%%%%%%%%%%%%%%%%%%%%%%%%%%%%%%%%%%%%%%%%%%%%%%%%%%%%%%%%%%%%%%%%%%%%%%%

\appendix

%%%%%%%%%%%%%%%%%%%%%%%%%%%%%%%%%%%%%%%%%%%%%%%%%%%%%%%%%%%%%%%%%%%%%%%%
% APPENDIX A
%%%%%%%%%%%%%%%%%%%%%%%%%%%%%%%%%%%%%%%%%%%%%%%%%%%%%%%%%%%%%%%%%%%%%%%%

\section{Uniqueness of ground states with $N=pN_c$ where $p$
is an integer}

Here we prove the uniqueness of those exact ground states
constructed in this paper for which the electron number $N$ is an
integer multiple of the number of unit cells $N_c$. The proof
proceeds as discussed in Sec. 2.3 and is exemplified by proving the
uniqueness of the ground state $|\Psi^{II}_g(4N_c)\rangle$,
(\ref{dia18}), of the diamond Hubbard chain. In this case $N=4N_c$,
i.e., $p=4$, and the electron density has the value $n=4/3$. The
Hamiltonian, the unnormalized ground state, and the $\hat A_{{\bf
i},\sigma}$ operators entering therein are given by (\ref{dia13}),
(\ref{dia18}),  and  (\ref{dia10}), respectively. To prove the
uniqueness one first has to determine the kernel of $\hat{H}^{\prime
}=\hat{H} -E^{II}_g
=\sum_{\sigma }\sum_{\mathbf{i}=1}^{N_{c}}\hat{A}_{\mathbf{i%
},\sigma }\hat{A}_{\mathbf{i},\sigma }^{\dagger }+U\hat{P}$. The
form of the Hamiltonian $\hat{H}^{\prime }$ implies that its kernel
is simply given by the intersection of the kernels of the individual
terms, i.e., $\ker (\hat{H}^{\prime})=\ker
(\sum_{\sigma}\sum_{\mathbf{i}=1}^{
N_{c}}\hat{A}_{\mathbf{i},\sigma}\hat{A}_{\mathbf{i},\sigma}^{\dagger})
\bigcap\ker (\hat{P})$, where $\hat{P}$ is an operator, which
assumes its minimum eigenvalue 0, if there is at least one electron
at each site.

To prove the uniqueness of the ground state
$|\Psi^{II}_g(4N_c)\rangle$, (\ref{dia18}), it is now necessary to
demonstrate that it spans $ker(\hat H')$. According to the
discussion in Sec. 2.3 the proof consists of two parts. First one
has to establish that

(i) $|\Psi^{II}_g(4N_c)\rangle$ is an element of the intersection of
the kernels of the positive semidefinite operators entering in $\hat
H'$.

Then one must show that

(ii) all states from the kernel of $\hat H'$ can be written in the
form of the ground state $|\Psi^{II}_g(4N_c)\rangle$.

The first part of the proof, (i), is covered by the following
Theorems A-1 to A-3. The second part of the proof, (ii), then
follows immediately (see below).

To prove (i) we note  that the first term in $\hat{H}^{\prime }$ is
a sum of positive semidefinite operators $\hat{A}_{\mathbf{i},\sigma
}\hat{A}_{\mathbf{i},\sigma }^{\dagger }$. The ground state of
$\hat{H}^{\prime}$ must therefore be an element of each kernel of
all the operator
$\hat{A}_{\mathbf{i},\sigma}\hat{A}_{\mathbf{i},\sigma}^{ \dagger}$,
for all $\mathbf{i},\sigma $. Thus we first determine the kernel of
$\hat{A}_{\mathbf{i},\sigma }\hat{A}_{\mathbf{i},\sigma }^{\dagger
}$ (Sec. A.1.), then the kernel of the sum $\sum_{\sigma
}\sum_{\mathbf{i}=1}^{N_{c}}\hat{A}_{\mathbf{i} ,\sigma
}\hat{A}_{\mathbf{i},\sigma }^{\dagger }$ (Sec. A.2.), and finally
the kernel of the full Hamiltonian $\hat{H}^{\prime}$ (Sec. A.3.).

\subsection{The kernel of $\mbox{\boldmath$\hat{A}_{i,\protect\sigma}
\hat{A}_{i,\protect\sigma}^{\dagger}$}$}

\textbf{Theorem A-1}: The kernel of $\hat{A}_{\mathbf{i},\sigma }\hat{A}_{%
\mathbf{i},\sigma }^{\dagger }$ is spanned by states of the form
\begin{eqnarray}
|\Psi \rangle =\hat{A}_{\mathbf{i},\sigma
}^{\dagger}\hat{W}^{\dagger }|0\rangle ,  \label{v}
\end{eqnarray}
where $\hat{W}^{\dagger }$ is an arbitrary operator.

\noindent\textit{Proof of Theorem A-1}

\noindent a) From ($\hat{A}_{\mathbf{i},\sigma}^{\dagger})^{2}=0$ it
follows that $\hat{A}_{\mathbf{i},\sigma }\hat{A}_{\mathbf{i},\sigma
}^{\dagger }|\Psi \rangle =0$. This implies that
$|\Psi\rangle\in\ker (\hat{A}_{\mathbf{i}, \sigma
}\hat{A}_{\mathbf{i},\sigma }^{\dagger })$.

\noindent b) In order to show that all elements of $\ker
(\hat{A}_{\mathbf{i}, \sigma}\hat{A}_{\mathbf{i},\sigma}^{\dagger
})$ can be written in the form
$|\Psi\rangle=\hat{A}_{\mathbf{i},\sigma}^{\dagger}\hat{W}^{\dagger}|0\rangle$
we consider a state $|\Phi\rangle =\hat{Y}^{\dagger}|0\rangle$
$\in\ker (
\hat{A}_{\mathbf{i},\sigma}\hat{A}_{\mathbf{i},\sigma}^{\dagger })$
and show that $|\Phi\rangle$ has the form $|\Psi\rangle =\hat
A^{\dagger}_{ \mathbf{i},\sigma}\hat W^{\dagger}|0\rangle$.

Making use of
$\left\{\hat{A}_{\mathbf{i},\sigma},\hat{A}_{\mathbf{i},\sigma
}^{\dagger}\right\}=8t_{\perp}^{2}t_{\parallel}+2t_{\parallel}=:a_{\mathbf{i},
\mathbf{i}}^{\sigma }>0$ and
$\hat{A}_{\mathbf{i},\sigma}\hat{A}_{\mathbf{i},
\sigma}^{\dagger}|\Phi\rangle =0$ we can write
\begin{eqnarray}
|\Phi \rangle =\hat{Y}^{\dagger }|0\rangle
=\frac{1}{a_{\mathbf{i},\mathbf{i} }^{\sigma
}}(\hat{A}_{\mathbf{i},\sigma }\hat{A}_{\mathbf{i},\sigma }^{\dagger
}+\hat{A}_{\mathbf{i},\sigma }^{\dagger }\hat{A}_{\mathbf{i} ,\sigma
})\hat{Y}^{\dagger }|0\rangle =\hat{A}_{\mathbf{i},\sigma }^{\dagger
}(\frac{1}{a_{\mathbf{i},\mathbf{i}}^{\sigma
}}\hat{A}_{\mathbf{i},\sigma } \hat{Y}^{\dagger })|0\rangle .
\end{eqnarray}
With the identification $\hat{W}^{\dagger
}:=\frac{1}{a_{\mathbf{i},\mathbf{i }}^{\sigma
}}\hat{A}_{\mathbf{i},\sigma }\hat{Y}^{\dagger }$ we have therefore
verified that indeed
\begin{eqnarray}
|\Phi \rangle =\hat{A}_{\mathbf{i},\sigma }^{\dagger
}\hat{W}^{\dagger }|0\rangle =|\Psi \rangle.
\end{eqnarray}
This completes the proof of Theorem A-1.

\subsection{ The kernel of $\mbox{\boldmath$\sum_{\protect\sigma}\sum_{i=1}^{
N_{c}}\hat{A}_{i,\protect\sigma}\hat{A}_{i,\protect\sigma}^{\dagger}$}$}

The kernel of $\sum_{\sigma
}\sum_{\mathbf{i}=1}^{N_{c}}\hat{A}_{\mathbf{i} ,\sigma
}\hat{A}_{\mathbf{i},\sigma }^{\dagger }$ has the form
\begin{eqnarray}
\ker (\sum_{\sigma
}\sum_{\mathbf{i}=1}^{N_{c}}\hat{A}_{\mathbf{i},\sigma }
\hat{A}_{\mathbf{i},\sigma }^{\dagger })=\underset{\sigma =\uparrow
,\downarrow }{\bigcap
}\underset{\mathbf{i}=1}{\overset{N_{c}}{\bigcap }} \ker
(\hat{A}_{\mathbf{i},\sigma }\hat{A}_{\mathbf{i},\sigma }^{\dagger
}).
\end{eqnarray}

\textbf{Theorem A-2}: $\ker (\sum_{\sigma
}\sum_{\mathbf{i}=1}^{N_{c}}\hat{A}_{ \mathbf{i},\sigma
}\hat{A}_{\mathbf{i},\sigma }^{\dagger })$ is spanned by vectors of
the form
\begin{eqnarray}
|\Psi \rangle =[\prod_{\sigma }\prod_{\mathbf{i}
=1}^{N_{c}}\hat{A}_{\mathbf{i},\sigma }^{\dagger }]\hat{W}^{\dagger
}|0\rangle,
\label{xxx1}
\end{eqnarray}
where the operator $\hat{W}^{\dagger }$ is arbitrary.

Theorem A-2 follows directly from the linear independence of the
operators $\hat{A}_{\mathbf{i},\sigma}^{\dagger}$(that is,
$\hat{A}_{\mathbf{i},\sigma
}^{\dagger}\neq\sum_{\mathbf{j}\neq\mathbf{i}}\alpha_{\mathbf{j},\sigma}
\hat{A}_{\mathbf{j},\sigma}^{\dagger}$ holds for arbitrary
coefficients $\alpha_{\mathbf{j},\sigma}$ due to the fact that the
operators $\hat{c}_{
\mathbf{i}+\mathbf{r}_{1},\sigma}^{\dagger},\hat{c}_{\mathbf{i}+\mathbf{r}
_{2},\sigma}^{\dagger}$ enter only in
$\hat{A}_{\mathbf{i},\sigma}^{\dagger}$ and not in
$\hat{A}_{\mathbf{j},\sigma}^{\dagger}$, for $\mathbf{j}\neq
\mathbf{i}$.

\subsection{The kernel of $\mbox{\boldmath$\hat{H}^{\prime}$}$}

The kernel of $\hat{H}^{\prime }$ is given by $\ker
(\hat{H}^{\prime})=\ker (\sum_{\sigma}\sum_{\mathbf{i}=1}^{
N_{c}}\hat{A}_{\mathbf{i},\sigma}\hat{A}_{\mathbf{i},\sigma}^{\dagger})
\bigcap\ker (\hat{P})$, where $\hat{P}$ is an operator, which
assumes its minimum eigenvalue 0, if there is at least one electron
at each site. Hence, $\ker (\hat{H}^{\prime})$ is spanned by vectors
of the form in Eq. (\ref{v}), where now the operator
$\hat{W}^{\dagger}$ is no longer arbitrary, but must have a
\textit{specific} form to satisfy the abovementioned requirement
imposed by $\ker (\hat{P})$.

\textbf{Theorem A-3}: For $N=4N_{c}$ the operator
$\hat{W}^{\dagger}$ is given by
$\hat{W}^{\dagger}=\hat{F}_{\sigma}^{\dagger}$, where
$\hat{F}_{\sigma}^{ \dagger }$ creates two electrons with fixed spin
$\sigma$ in each unit cell.

\noindent\textit{Proof of Theorem A-3.}

The proof proceeds in two main steps: we show that
$\hat{W}^{\dagger}=\hat{F}_{ \sigma}^{\dagger}$ is 1) a possible
choice, and 2) the unique choice.

1) Let us denote by $\mathbf{T}_{\mathbf{i}}$ the unit cell (=
triangle), which contains the sites
$\mathbf{i},\mathbf{i}+\mathbf{r}_{1},\mathbf{i}+ \mathbf{r}_{2}$.
The state $\hat{F}_{\sigma}^{\dagger}|0\rangle$ has one empty site
in each unit cell $\mathbf{T_{i}}$. We now analyze how these empty
sites are filled by the product
$\prod_{\mathbf{i}=1}^{N_{c}}\hat{A}_{\mathbf{i},
\sigma}^{\dagger}$. We first observe that
$\hat{A}_{\mathbf{i},\sigma}^{ \dagger}$ acts on all sites in
$\mathbf{T}_{\mathbf{i}}$, which implies that the empty site in
$\mathbf{T}_{\mathbf{i}}$ can indeed be filled by applying
$\hat{A}_{\mathbf{i},\sigma}^{\dagger}$. There are $N_{c}$ empty
sites and $N_{c}$ operators $\hat{A}_{\mathbf{i},\sigma}^{\dagger}$;
thus, the product
$\prod_{\mathbf{i}=1}^{N_{c}}\hat{A}_{\mathbf{i},\sigma}^{\dagger}$
inserts $N_{c}$ electrons with spin $\sigma$ into the system. Since
($\hat{c}_{ \mathbf{j},\sigma}^{\dagger})^{2}=0$, the state
$\prod_{\mathbf{i}=1}^{N_{c}}
\hat{A}_{\mathbf{i},\sigma}^{\dagger}\hat{F}_{\sigma}^{\dagger}|0\rangle
$ has \textit{one} $\sigma $-electron on all $3N_{c}$ sites of the
lattice. Hence, $\hat{F}_{\sigma}^{\dagger}$ is a possible choice
for $\hat{W}^{\dagger}$ in constructing the states contained in
$\ker (\hat{H^{\prime}})$.

2) We now flip the spin $\sigma$ of one electron in
$\hat{F}_{\sigma}^{ \dagger}$ into $-\sigma$, and denote the new
operator $\hat{W}^{\dagger}$ by $\hat{F^{\prime}}_{\sigma
}^{\dagger}$. In this case, a single unit cell, say
$\mathbf{T}_{\mathbf{i}_{o}}$, will have one empty site, one
electron with spin $\sigma$ on another site, and one electron with
spin $-\sigma$ on the third site. Then the operator
$\hat{A}_{\mathbf{i}_{o},\sigma}^{\dagger}$ can create one electron
with spin $\sigma $ on the site, which already contains the electron
with spin $-\sigma$, thereby creating a double occupancy. One empty
site will remain in $\mathbf{T}_{\mathbf{i}_{o}}$. For
$\mathbf{i}\neq \mathbf{i}_{o}$, the operators
$\hat{A}_{\mathbf{i},\sigma}^{\dagger}$ can introduce their electron
into the empty sites of $\mathbf{T}_{\mathbf{i}}$. Hence one empty
site can remain in the system. The state $|\chi\rangle =$
$\prod_{\mathbf{i}=1}^{N_{c}}\hat{A}_{\mathbf{i},\sigma}^{\dagger}\hat{F^{
\prime}}_{\sigma}^{\dagger}|0\rangle$ therefore has at least one
term
with an empty site and contains one $-\sigma$ electron. Acting now
with
$\prod_{\mathbf{i}=1}^{N_{c}}\hat{A}_{\mathbf{i},-\sigma}^{\dagger}$
on $|\chi \rangle$ adds $N_{c}$ electrons with spin $-\sigma$ on
$(3N_{c}-1)$ sites. The new state $|\Phi\rangle
=\prod_{\mathbf{i}=1}^{N_{c}}\hat{A}_{\mathbf{i},-
\sigma}^{\dagger}|\chi\rangle $ will therefore have at least one
term
with an empty site, and is thus not an element of $\ker (\hat{P})$
and $\ker (\hat{H})$. This shows that
$\hat{W}^{\dagger}=\hat{F}_{\sigma}^{\dagger}$ is the
\textit{unique} choice. This completes the proof of Theorem A-3.

Theorems A-1, A-2 and A-3 prove that $|\Psi^{II}_g(4N_c)\rangle$,
(\ref{dia18}), is indeed an element of the intersection of the
kernels of the positive semidefinite operators entering in $\hat
H'$. This completes  part (i) of the proof of the uniqueness
discussed in Sec. 2.3.

Since all vectors from $\ker (\hat{H'})$ can be written in the form
$|\Psi^{II}_g(4N_c)\rangle$, (\ref{dia18}), and since this state
represents the constructed ground state $|\Psi_g\rangle$, it follows
immediately that \emph{all} vectors in $\ker (\hat{H}^{\prime})$ can
be written in the form (\ref{dia18}). Therefore the ground state
(\ref{dia18}) is the \emph{unique} ground state of the Hamiltonian
$\hat{H}$, (\ref{dia13}).

The proofs of the uniqueness of the ground state (\ref{dia5}) at
density $n=1/3$,  corresponding to $p=1$, and of solution III in
Ref. \cite{re15}  at density $n=5/3$ ($p=5$) proceed similarly.

%%%%%%%%%%%%%%%%%%%%%%%%%%%%%%%%%%%%%%%%%%%%%%%%%%%%%%%%%%%%%%%%%%%%%%%%%
% END APPENDIX A
%%%%%%%%%%%%%%%%%%%%%%%%%%%%%%%%%%%%%%%%%%%%%%%%%%%%%%%%%%%%%%%%%%%%%%%%%

%%%%%%%%%%%%%%%%%%%%%%%%%%%%%%%%%%%%%%%%%%%%%%%%%%%%%%%%%%%%%%%%%%%%%%%%%
% APPENDIX B
%%%%%%%%%%%%%%%%%%%%%%%%%%%%%%%%%%%%%%%%%%%%%%%%%%%%%%%%%%%%%%%%%%%%%%%%%

\section{Uniqueness of ground-state solutions with $\mbox{\boldmath$N=pN_c$}$
for non-integer $\mbox{\boldmath$p$}$}

\subsection{Geometric degeneracies}

For non-integer $p$ as in the case of (\ref{dia5}) for $n < 1/3$,
and (\ref{dia20}) for $n> 4/3$ (as well as in the case of solution
III for $n> 5/3$ of Ref.[\cite{re15}]), the ground states have a
high geometric degeneracy. In these cases there exist linearly
independent eigenstates $|\Psi_g^{\gamma}\rangle$ with the same
ground-state energy. They may be denoted by a degeneracy index
$\gamma =1,2, \ldots,M_d$, where $M_d$ is the degree of degeneracy.
The ground state then becomes $|\Psi_g\rangle =
\sum_{\gamma=1}^{M_d} \alpha_{\gamma} \: |\Psi_g^{\gamma}\rangle,
\label{degeq1}$ where apart from the normalization condition, the
coefficients $\alpha_{\gamma}$ are arbitrary. For these ground
states the proof of uniqueness proceeds analogously to that
described in Appendix A. Namely, one has to prove that $|\Psi_g
\rangle$ spans ker($\hat H'$), but now, given the specific structure
of $|\Psi_g\rangle$, it has to be verified that

\noindent 1) \emph{all} linearly independent states
$\left\vert\Psi_{g}^{\gamma} \right\rangle$ are elements of the
intersection of the kernels of the positive semidefinite operators
$\hat{P}_{n}$, i.e., $\left\vert\Psi _{g}^{\gamma} \right\rangle\in
{\ker (\hat{H}^{\prime})=\overset{L}{\underset{n=1}{\bigcap}} \ker
(\hat{P}_{n})}$, and

\noindent 2) \textit{all} states $\left\vert\Psi
_{g}\right\rangle\in { \overset{L}{\underset{n=1}{\bigcap}}\ker
(\hat{P}_{n})}$ can be written in the form of the constructed ground
state, e.g. as a linear combination of the vectors
$|\Psi_g^{\gamma}\rangle$.

As an example we now prove the uniqueness of the ground state
$|\Psi^{I}_g(N)\rangle$, (\ref{dia5}), for $n<1/3$. The proof is
presented for fixed $N$, namely $N=N_c-1$. For other $N$ values the
proof proceeds similarly.
 The first part of the proof, (i), (see Sec. 2.3) is covered by the following
 Theorems B-1 and B-2. The second part of the proof, (ii), then
 follows immediately (see below).

\subsection{Uniqueness of the ground state (\ref{dia5}) for
$\mbox{\boldmath$N=N_c -1$}$ particles}

The transformed Hamiltonian $\hat H=\hat H_0+\hat H_U$ is determined
by $\hat H_0$ and $\hat H_U$ as given by (\ref{dia4})and
(\ref{dia1}), respectively. Furthermore, one has $\hat N= \sum_{{\bf
i},\sigma} \sum_{\nu=1}^3 \hat C^{\dagger}_{\nu,{\bf i},\sigma} \hat
C_{\nu,{\bf i},\sigma}$, where $\hat C_{\nu,{\bf i},\sigma}$ are
genuine canonical Fermi operators, and $E_3 < E_1,E_2$. Since $\hat
N$ is a constant of motion, the Hamiltonian can be written as $\hat
H= \sum_{{\bf i},\sigma} [ (E_{1}-E_3) \hat C^{\dagger}_{1,{\bf
i},\sigma} \hat C_{1,{\bf i},\sigma} + (E_{2}-E_3) \hat
C^{\dagger}_{2,{\bf i},\sigma} \hat C_{2,{\bf i},\sigma} ] + \hat
H_U + E^{I}_g$, where $ E^{I}_g=E_{3}N$ and $(E_1-E_3),(E_2-E_3)
> 0$. Consequently, $\hat H' := \hat H- E^{I}_g$ is a positive
semidefinite operator: $\hat H'=\hat P_1+\hat P_2$, where $\hat
P_1=\sum_{{\bf i},\sigma} [(E_{1}-E_3)\hat C^{\dagger}_{1,{\bf
i},\sigma} \hat C_{1,{\bf i},\sigma} + (E_{2}-E_3) \hat
C^{\dagger}_{2,{\bf i},\sigma} \hat C_{2,{\bf i},\sigma}]$, and
$\hat P_2=\hat H_U$. The ground state is given by (\ref{dia5}).

\subsubsection{The kernel of $\hat P_1$}

{\bf Theorem B-1}: The kernel of $\hat P_1$ is spanned by states of
the form
\begin{eqnarray}
|\Psi(\{ {\bf i} \}, \{\sigma_{\bf i} \}) \rangle = \prod_{{\bf
i}=1}^{N_c -1} \hat C^{\dagger}_{3,{\bf i}, \sigma_{\bf i}}| 0
\rangle , \label{degeq5}
\end{eqnarray}
where the product is taken over arbitrary $N_c-1$ sites, and
$\sigma_{\bf i}$ is arbitrary. Note that $|\Psi(\{{\bf i}
\},\{\sigma_{\bf i} \})\rangle$ depends on the set of sites $\{ {\bf
i}\}$, and the set of spin orientations $\{\sigma_{\bf i}\}$.

\noindent\textit{Proof of Theorem B-1.}

\noindent The proof proceeds in two steps: a) one first shows that
Eq. (\ref{degeq5}) is a possible choice, and b) then verifies that
it is the unique choice.

a) The initial $6 N_c$ canonical Fermi operators $\hat c_{{\bf
i}+{\bf r}_s, \sigma}$, where $s=1,2,3$ is the sublattice index,
have been transformed by a linear transformation into the $6 N_c$
opertors $\hat C_{\nu,{\bf i},\sigma_{ \bf i}}$, which satisfy
canonical fermionic anti-commutation rules. Since the set of the new
operators $\{\hat C_{\nu,{\bf i},\sigma_{\bf i}} \}$ substitutes the
set $\{\hat c_{{\bf i}+{\bf r}_s,\sigma} \}$, all operators in the
system acting on states defined by arbitrary $({\bf i},\sigma_{\bf
i})$ can be expressed in terms of the operators $\hat C_{\nu,{\bf
i},\sigma_{\bf i} }$.

Given these conditions, an operator $\hat X^{\dagger}$ which
satisfies
\begin{eqnarray}
\{\hat C_{1,{\bf i},\sigma_{\bf i}}, \hat X^{\dagger} \} = \{\hat
C_{2,{\bf i},\sigma_{\bf i}}, \hat X^{\dagger} \} = 0
\label{degeq5a}
\end{eqnarray}
for all indices ${\bf i},\sigma_{\bf i}$, must have the form $\hat
X^{\dagger}=\prod_{\bf j}\prod_{ \sigma_{ \bf j}}\hat
C^{\dagger}_{3,{\bf j},\sigma_{\bf j}}$. Indeed, the
anti-commutation relations (\ref{degeq5a}) hold only for this type
of   $\hat X^{\dagger}$ operator since an arbitrary $\hat
X^{\dagger}$ operator can be expressed in terms of $\hat C_{\nu,{\bf
j},\sigma_{\bf j}}$ operators, and only $\hat C^{\dagger}_{3,{\bf
j},\sigma_{\bf j}}$ anticommutes with all $\hat C_{1,{\bf
i},\sigma_{\bf i}}$ and $\hat C_{2,{\bf i},\sigma_{\bf i}}$.

b) If $\hat X^{\dagger}$ contains at least one $\hat
C^{\dagger}_{\nu,{\bf j}, \sigma_{\bf j}}$ operator with $\nu=1,2$,
the relation Eq. (\ref{degeq5a}) no longer holds, hence Eq.
(\ref{degeq5}) is the unique choice. This proves Theorem B.1.

We furthermore observe that due to the fermionic operator nature of
$\hat C_{3, {\bf i},\sigma_{\bf i}}$, the deduced elements of
ker($\hat P_1$) from Eq. (\ref{degeq5}) are not only linearly
independent but also orthogonal:
\begin{eqnarray}
\langle\Psi(\{ {\bf i} \},\{\sigma_{\bf i} \}) |\Psi(\{ {\bf
i}'\},\{\sigma_{ \bf i}' \})\rangle =\delta_{\{ {\bf i} \},\{ {\bf
i}' \}} \delta_{\{\sigma_{\bf i} \},\{\sigma_{\bf i}' \} } .
\label{degeq6}
\end{eqnarray}

\subsubsection{ The kernel of $\hat H'$}

Since ker($\hat H'$)=ker($\hat P_1$) $\cap$ ker($\hat P_2$), the
states in ker($\hat H'$) have the form (\ref{degeq5}), but the
specific set of indices $(\{ {\bf i} \},\{\sigma_{\bf i} \})$ has to
be chosen such that these states are also contained in ker($\hat
P_2$). This condition requires the proof of the following theorem.

{\bf Theorem B-2}: The kernel of $\hat H'$ is spanned by states of
the form
\begin{eqnarray}
|\Psi ({\bf j})\rangle = |\Psi(\{ {\bf i} \}) \rangle = \prod_{{\bf
i}=1}^{N_c -1} \hat C^{\dagger}_{3,{\bf i},\sigma}| 0 \rangle ,
\label{degeq7}
\end{eqnarray}
where $\sigma$ is fixed, and the product extends over $N_c-1$
arbitrary sites.

Since the product in Eq. (\ref{degeq7}) is taken over $N_c-1$ sites
of the sublattice $s=3$, the vector $|\Psi(\{ {\bf i} \})\rangle$
can be denoted by the specific site ${\bf j}$ excluded from the
product, i.e. $|\Psi(\{ {\bf i} \})\rangle\equiv |\Psi ({\bf
j})\rangle$. This site ${\bf j}$ plays the role of the degeneracy
index $\gamma$ in Sec.B.1.

\textit{Proof of Theorem B-2.}

\noindent The theorem is proved in three steps: First one has to
show that all $\sigma_{\bf i}$ spin indices in (\ref{degeq7}) must
have the same value, secondly it has to be verified that the deduced
states in ker($\hat H'$) are linearly independent, and thirdly one
needs to prove that the set of the deduced states is complete.

a) All states in ker($\hat P_2$) require the absence of any double
occupancy. Therefore, all indices $\sigma_{\bf i}$ must be fixed to
the same value $\sigma$. The double occupancy cannot be avoided
otherwise, since in $N_c$ unit cells $N_c-1$ operators $\hat
C^{\dagger}_{3,{\bf i},\sigma}$ share a common site.

b) due to the relation (\ref{degeq6}) the vectors $|\Psi({\bf
j})\rangle$ are linearly independent, since they are orthogonal
$\langle \Psi({\bf j}) | \Psi({\bf j}') \rangle = \delta_{{\bf j},
{\bf j}'}$.

c) Obviously, the degree of degeneracy of a macroscopic many-body
state is given by the number of equivalent configurations. Due to
part a), apart from the trivial degeneracy with respect to the total
spin direction, on finds $M_d=N_c$, because there are $N_c$
possibilities for placing $N_c-1$ identical objects on $N_c$ sites.
Consequently, ker($\hat H'$) is an $M_d=N_c$ dimensional Hilbert
subspace. The index ${\bf j}$ in the states of the form
(\ref{degeq7}) refers to $N_c$ possible different unit cells.
Furthermore, the vectors $|\Psi({\bf j})\rangle$ are linearly
independent, hence the set $\{|\Psi({\bf j})\rangle \}$ forms a
(orthonormalized) basis of ker($\hat H'$). Therefore, $\{|\Psi({\bf
j})\rangle \}$ spans ker($\hat H'$), and the ground state has the
general form $|\Psi_g\rangle = \sum_{{\bf j}=1}^{N_c} \alpha_{\bf j}
|\Psi({\bf j})\rangle$, with arbitrary coefficients $\alpha_{\bf
j}$. This proves Theorem B-2.

Since $\{|\Psi({\bf j})\rangle \}$ is a basis of ker($\hat H'$),
\emph{all} states $|\phi\rangle \in ker(\hat H')$ can be written in
the form $|\Psi_{g}\rangle$ of part c). Therefore (\ref{dia5}) is
the unique ground state of the Hamiltonian $\hat H$ consisting of
the two terms (\ref{dia4}) and (\ref{dia1}).

For $N=N_c$ the set $\{ {\bf j} \}$ is an empty set, and the upper
limit set to $N_c$ in Eq. (\ref{degeq7}) provides the unique ground
state at density $n=1/3$ (see (\ref{dia5}) for $N=N_c$). A similar
strategy for $N < N_c -1$ allows one to prove the uniqueness of the
ground-state solutions for all electron densities $n < 1/3$ in
(\ref{dia5}). Similar proofs are expected to hold in the case of the
ground state $|\Psi^{II}_g (4N_c + \Delta N)\rangle$, (\ref{dia20}),
for $n>4/3$ ($p>4$) and solution III for $n>5/3$ ($p>5$) constructed
in Ref. \cite{re15}.

%%%%%%%%%%%%%%%%%%%%%%%%%%%%%%%%%%%%%%%%%%%%%%%%%%%%%%%%%%%%%%%%%%%%%%%%%%%%%
% END APPENDIX B.
%%%%%%%%%%%%%%%%%%%%%%%%%%%%%%%%%%%%%%%%%%%%%%%%%%%%%%%%%%%%%%%%%%%%%%%%%%%%%

\end{document}